\documentclass{iaus}
\usepackage{graphicx}
\def\fermilat{\textit{Fermi}/LAT}

\title[IAU Symposium 275: Jets at all Scales] 
{The jet in M87 from e-EVN observations}

\author[G. Giovannini et al.]
{G. Giovannini$^{1,2}$, C. Casadio$^1$, M. Giroletti$^1$, 
M. Beilicke$^3$, A. Cesarini$^4$ \& H. Krawczynski$^3$}   

\affiliation{$^1$ Istituto di Radioastronomia-INAF, \\ via Gobetti 101, 40129 
Bologna, Italy \\ email: {ggiovann@ira.inaf.it} \\
[\affilskip]
$^2$ Dipartimento di Astronomia, \\ via Ranzani 1, 40127 Bologna, Italy \\
[\affilskip]
$^3$ Department of Physics, Washington University, \\ St. Louis, MO 63130, 
USA \\
[\affilskip]
$^4$ School of Physics, National University of Ireland \\
Galway, University Road, Galway, Republic of Ireland}

\pubyear{2010}
\volume{275}  
\pagerange{119--126}
\jname{Jets at all Scales}
\editors{G.E. Romero, R.A. Sunyaev\& T. Belloni, eds.}
\begin{document}

\maketitle

\begin{abstract}
One of the most intriguing open questions of today's astrophysics
     is the jet physical properties and the location and the mechanisms for 
the production
     of MeV, GeV, and TeV gamma-rays in AGN jets. M87 is a privileged
     laboratory for a detailed study of the properties of jets, owing to its
     proximity, its massive black hole, and its conspicuous emission at radio
     wavelengths and above. We started on November 2009 a monitoring program
     with the e-EVN at 5 GHz. 
We present here results of these multi-epoch
     observations and discuss the two episodes of activity at energy
     E$>$100 GeV that occured in this period. 
One of these observations was obtained at the same day of
     the first high energy flare. 
We added to our results literature data obtained with the VLBI and VLA.
A clear change in the proper motion velocity
     of HST-1 is present at the epoch $\sim$ 2005.5. In the time range
     1998 -- 2005.5 the apparent velocity is subluminal, and superluminal
($\sim$ 2.7c) after 2005.5.
\end{abstract}

\keywords{Galaxies: jets, Galaxies: M87, Galaxies: active}

\firstsection 
\section{Introduction}

The giant radio galaxy Messier 87 (M87), also known as 3C 274 or Virgo A, is
one of the best studied radio sources and a known
$\gamma$-ray-emitting AGN.  It
is located at the center of the Virgo cluster of galaxies at a distance = 16.7
Mpc, corresponding to an angular conversion 1 mas = 0.081 pc. The massive black
hole at the M87 center has an estimated mass = 6 $\times$ 10$^9$ solar masses,
with a scale of 1 mas = 140 R$_S$. The bright jet is well resolved in the
X-ray, optical, and radio wave bands.

\begin{figure}[b]
\begin{minipage}{14pc}
 \includegraphics[width=11pc,angle=-90]{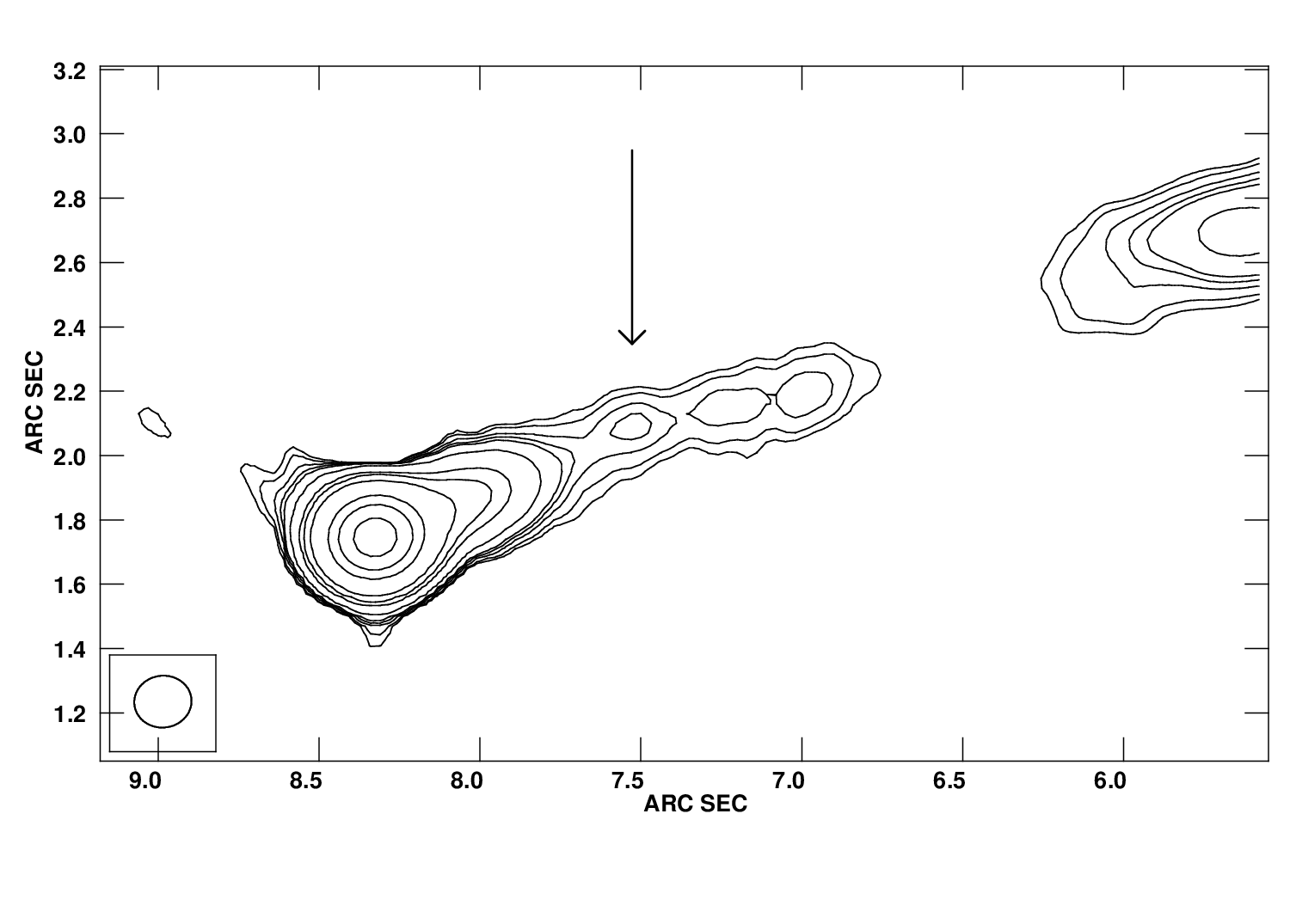}
 \caption{M87 jet obtained on March 1998 at 15 GHz with the VLA in A 
configuration.
An arrow indicates the HST-1 position. Levs are: -3 1.5 2 3 
4 5 10 30 50 100 500 1000 2000 mJy/beam. The HPBW is 0.16''}
   \label{fig1}
\end{minipage}
\hspace{2pc}
\begin{minipage}{14pc}
 \includegraphics[width=11pc,angle=-90]{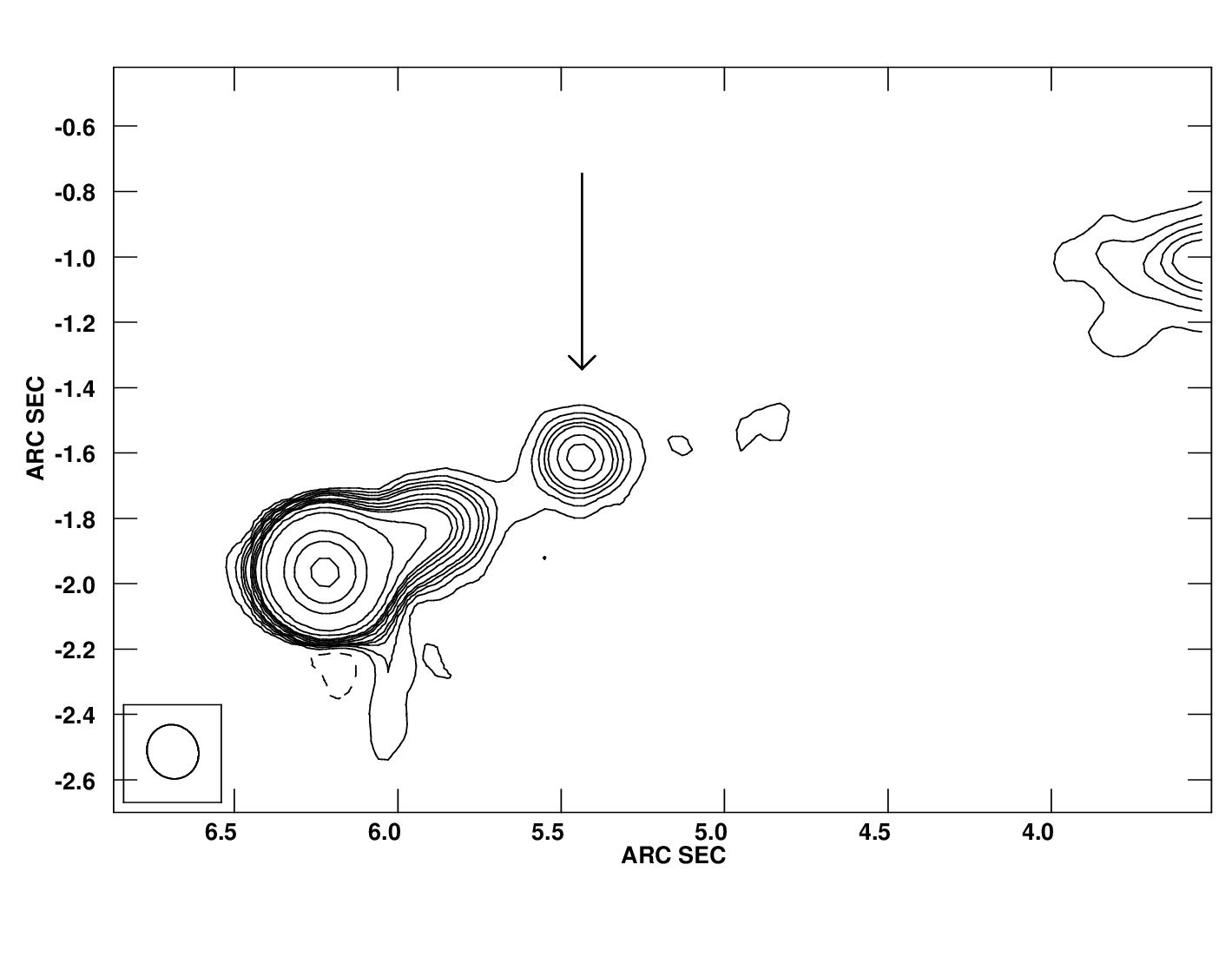}
 \caption{M87 jet obtained on June 2003 at 15 GHz with the VLA in A 
configuration.
An arrow indicates the HST-1 position. Levs are: -3 2 
4 6 8 10 15 20 30 50 100 500 1000 2000 mJy/beam. The HPBW is 0.16''}
   \label{fig2}
\end{minipage}
\end{figure}

The jet is characterized by many substructures and knots.  In 1999 HST
observations revealed a bright knot at about 1'' from the core, named
HST-1. This feature is active in the radio, optical, and X-ray regimes.  It was
discussed by \cite[Perlman et al. 1999]{Perlman1999}, who compared optical 
and radio images.
\cite[Biretta et al. 1999]{Biretta1999} measured in the range 1994-1998 a 
subluminal speed = 0.84c
for the brightest structure (HST-1 East), which appears to emit superluminal
features moving at 6c. However this motion was measured in regions on a
larger scale with respect to the VLBI structures discussed here.
In this time range HST-1 in the radio band was a faint jet structure (a few 
mJy/beam, see Fig.\,\ref{fig1}), but starting from 2000 it increased by more
than a factor 50 and it reached a flux density $\sim$ 100 mJy in 2005 
(See Fig.\,\ref{fig2}
and \cite[Harris et al. 2009]{Harris2009}). 

VLBI observations of the M87 inner region show a well resolved, edge-brightened
jet structure. At very high resolution (43 and 86 GHz) near to the brightest
region the jet has a wide opening angle, and we refer to the many published
papers which discuss the possible presence of a counter-jet and the location of
the radio core; see e.g. \cite[Junor et al. 1999]{Junor1999}, 
\cite[Krichbaum et al. 2005]{Krichbaum2005}. After a few milliarcsec (mas) 
the jet appears well collimated and limb-brightened.

Very High Energy (VHE) $\gamma$-ray emission was reported by the High Energy
Gamma-Ray Astronomy (HEGRA) collaboration in 1998/99 
(\cite[Aharonian et al. 2003]{Aharonian2003}),
confirmed by the High Energy Stereoscopic System (HESS) in 2003-2006
(\cite[Aharonian et al. 2006]{Aharonian2006}), and by VERITAS in 2007 
(\cite[Acciari et al. 2008]{Acciari2008}). Coordinated
intensive campaigns have permitted to detect the source again in 2008
(\cite[Acciari et al. 2009]{Acciari2009}) and as recently as February and 
April 2010
(\cite[Mariotti et al. 2010]{Mariotti2010}). 
Steady emission at MeV/GeV energies has also
been detected by \fermilat ~~(\cite[Abdo et al. 2009]{Abdo2009}).

Various models have been proposed to explain the multi-wavelength emission and
in particular to constrain the site of the VHE emission in M87. The
inner jet region was favoured by the observed short TeV variability timescales
according to \cite[Aharonian et al. 2006]{Aharonian2006}. The VHE emission 
could then be 
produced in
the BH magnetosphere (\cite[Neronov et al. 2007]{Neronov2007}) or in the 
slower 
jet layer
(\cite[Tavecchio et al. 2008]{Tavecchio2008}), with the spine accounting for 
the emission from
the radio to the GeV band; this would lead to a complex correlation between the
TeV and radio components.

However, VLBA observations at 1.7 GHz by \cite[Cheung et al. 2007]{Cheung2007} 
resolved HST-1 in
substructures with superluminal components.  
\cite[Aharonian et al. 2006]{Aharonian2006}
discussed that HST-1 cannot be excluded as a source of TeV $\gamma$ rays,
however they conclude that the more promising possibility is that the site of
TeV $\gamma$-ray production is the nucleus of M87 itself. Comparing
multifrequency data \cite[Harris et al. 2008]{Harris2008} suggested that the 
TeV emission from M87 was originated in HST-1.

Finally, \cite[Acciari et al. 2009]{Acciari2009} reported rapid TeV flares 
from M87 in February
2008, associated by an in increase of the radio flux from the nucleus, while
HST-1 was in a low state, thus concluding that the TeV flares originate in the
core region.

In this context we started at the end of 2009 a program to observe with the
e-EVN
M87 at 5 GHz to study the properties of the M87 core, jet, and HST-1 structure.

\section{Observations and Data Reduction}

The observations have been carried out in e-VLBI mode, with data acquired by
EVN
radio telescopes, directly streamed to the central data processor at JIVE, and
correlated in real-time. The observing frequency of 5 GHz was chosen to
simultaneously grant a large field of view and a high angular resolution. For
observations taking advantage of the long baselines provided by the Arecibo and
Shanghai telescopes, our clean beam with uniform weights is about 
$2.0 \times 0.9$ mas in PA $-25^\circ$.

We obtained 6 epochs at 5 GHz, namely on 2009 November 19, 2010 
January 27, February 10, and March 28, and as Target of Opportunity on 
2010 March 6, May 18, and June 9.

As a result of the large bandwidth (a rate of 1 Gbps was sustained by most
stations),
long exposure (up to 6 hours per epoch), and extended collecting area, the rms
noise in our images is mostly dynamic range limited. As an average value, we
can quote 0.5 -- 0.8 mJy~beam$^{-1}$ in the nuclear region and 0.1 -- 0.2
mJy~beam$^{-1}$ in the HST-1 region.
We present here preliminary results. Data reduction was carried out in the
standard mode using the AIPS and CalTech package.

\section{Results}

\subsection{The inner jet region}

The jet orientation and velocity has been discussed
in many papers comparing observational data on the jet brightness and
proper motion. Recently \cite[Acciari et al. 2009]{Acciari2009} assumed as a 
likely range $\theta$ = 15 -- 25 deg.

Because of the very similar uv-coverage, we used our images to search for 
evidence of a
possible proper motion, comparing different epoch position of jet
substructures and subtracting images at different epochs (with the
same grid, angular resolution and similar uv-coverage) to look for
possible systematic trends. No evidence was found in anycase.

\begin{figure}[b]
\begin{center}
 \includegraphics[width=3.8in]{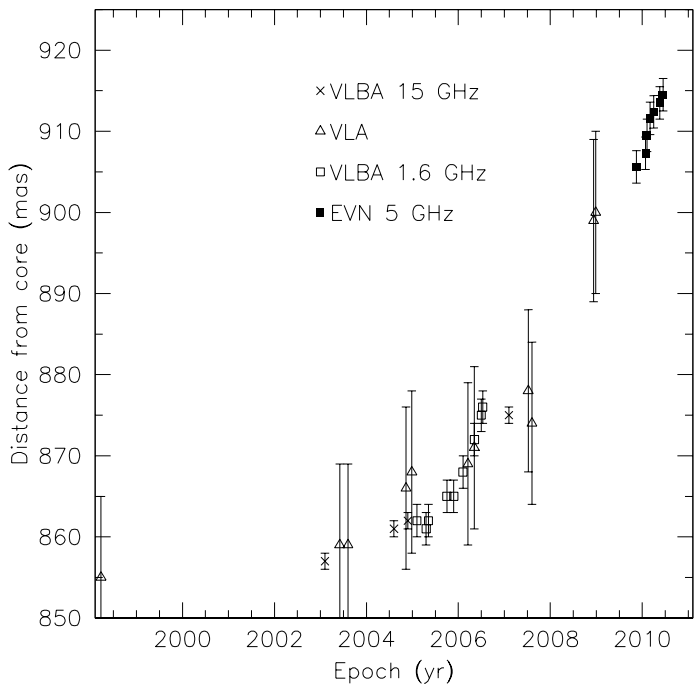}
 \caption{Distance of HST-1 brightest peak from the M87 core at
      different epochs}
   \label{fig4}
\end{center}
\end{figure}

We find a
marginal evidence of a nuclear flux density increasing in the last three 
epochs. In these epochs we have an increase of the core flux density and
of the inner jet (within $\sim$ 8 mas) flux density. The data analysis
is still in progress since it is not easy to separate the core and jet flux 
density because of the source structure.

\subsection{HST-1}

In our observations HST-1 is clearly resolved. 
It is oriented in E-W direction forming an angle of $\sim$ 20$^\circ$ with the
jet axis. The HST-1 size is in agreement with a very small ($\sim$ 0$^\circ$)
jet opening angle confirming the high jet collimation in the sub-arcsecond 
region.


To better study the dynamic of this structure we searched archive VLA data at
high resolution (A configuration) and high frequency (U, and Q bands).
We refer to \cite[Harris et al. 2009]{Harris2009} for a discussion of the 
flux density
variability. Here we only want to compare different epochs to derive the HST-1
dynamic.

We started to analyze data from 1998, even if HST-1 is very faint before of 
2003.
Starting from 2003.6 the HST-1 structure
is well evident (see e.g. Fig.~\ref{fig2} obtained on June 2003) and
well separated by the jet structure near the
core.

We estimated from e-EVN and VLA data the distance of HST-1 from the core.
In e-EVN data we measured the distance between the core and the brightest
knot in HST-1, in VLA images we used the HST-1 peak, being
this structure unresolved. Adding the values obtained at 1.5 and 15 GHz
by \cite[Cheung et al. 2007]{Cheung2007} and 
\cite[Chang et al. 2010]{Chang2010}, respectively,
we can study the HST-1 proper motion with a good statistic from 2003 to
present epoch. The apparent proper motion of HST-1 is shown in
Fig.~\ref{fig4}.

A clear change in the proper motion velocity is present at the epoch
$\sim$ 2005.5, coincident with the TeV $\gamma$-ray activity and
the maximum radio/X-ray flux density of
HST-1. In the time range 2003 -- 2005.5 the apparent velocity is
0.5c -- 0.6c; in the time range 2005.5 -- 2010.25 the apparent velocity is
$\sim$ 2.7c. We note also a possible decrease in the apparent velocity in 
2007 with
a restarted high velocity motion from 2008 (near the time of the high energy 
flare) up to now.
Assuming a jet orientation angle = 25$^\circ$ a proper motion of 2.7c
corresponds to an intrinsic velocity = 0.94c.

\section{Summary}

With our new e-EVN data 
we have obtained images of the nuclear region of M87 and of the jet
substructure HST-1.

The radio core flux density is constant in the first three epochs
with an average flux density $\sim$ 1805 mJy and slightly increasing in the
last three epochs: 2013 mJy in 2010.25.

The HST-1 structure is well resolved in many substructures. A 
complex proper motion
is clearly present. Comparing e-EVN data with archive VLA data and published
VLBA data at 1.7 and 15 GHz we find a strong evidence that in 2005.5
HST-1 increased its velocity from an apparent velocity $\sim$ 0.5c
to 2.7c. With present data it is not possible to discuss if this change in
velocity is related to the M87 VHE activity and/or to the maximum radio/X-ray
flux density of HST-1 at this
epoch. A more regular and longer
monitor and a multi-frequency comparison is necessary to clarify this point.


\end{document}